# Precise match for the new experimental fine structure constant value

E. M. Lipmanov
40 Wallingford Road # 272, Brighton MA 02135, USA

**Abstract**

In this note I point out a highly accurate relation - to within 0.2 ppb - between the recently discovered accurate experimental value of the fine structure constant $(1/\alpha)_{exp}$ =137.035999710(96) and the precise solution $(1/\alpha)$= 137.035999737 of a primary nonlinear equation for the fine structure constant generated by the earlier suggested new universal flavor-electroweak phenomenological constant $\alpha_o \equiv \exp(-5)$ as a source and hints from the new experimental data. This precise match for the experimental value $(1/\alpha)_{exp}$ lends additional support to the notion of a new universal dimensionless physical constant $\alpha_o$.

---

**1. Introduction.** In the recent paper by G. Gabrielse, D. Hanneke, T. Kinoshita, N. Nio and B. Odom [1] low energy value of the fine structure constant is determined by a new measurement of the electron (g-2)-factor in a single-electron quantum cyclotron experiment [2] and improved calculations in QED theory:

$$\alpha_{exp} = 1/137.035999710(96). \tag{1}$$

The small uncertainty of this result is ~0.7 ppb, about 10 times better than in any other method to determine the value of the fine structure constant $\alpha$.

Another estimation of the fine structure constant in ref.[3] with data inputs from the same experiment [2] is given by $\alpha_{exp}$ = 1/137.035999709(96), it is in exact agreement with the result (1) from [1].

And further improvements are in prospect [2, 1].

So, there is a challenge to match the very high accuracy of the experimental value (1) with theoretical or phenomenological estimation, especially if the estimation implies a relation of the fine structure constant to other basic physical quantities.

**2. Precise equation for the fine structure constant.** In this Note I consider an exact phenomenological match for the value $\alpha_{exp}$ (1) that is generated by the suggested earlier [4, 5] new dimensionless universal flavor-electroweak physical constant,

$$\alpha_o \equiv \exp(-5), \qquad (2)$$

as the source-value. The match is a solution of the following essentially nonlinear equation for the unknown $\alpha$,

$$(\exp\alpha\,/\alpha)^{\exp 2\alpha} + (\alpha/\pi)[1-(\alpha_o/\pi)] = 1/\alpha_o = \exp 5. \qquad (3)$$

Equation (3) is characterized by two special terms - one main exponential nonlinear in $\alpha$ first term and a much smaller linear in $(\alpha/\pi)$ second one.

The accurate numerical solution of equation (3) is given by

$$\alpha \cong 1/137.0359997426. \qquad (4)$$

A more compact form of Eq.(3),

$$(\exp\alpha\,/\alpha)^{\exp 2\alpha} + (\alpha/\pi)\exp(-\alpha_o/\pi) = 1/\alpha_o, \qquad (3')$$

leads to the equally accurate solution:

$$\alpha \cong 1/137.0359997372. \qquad (4')$$

The main result is that the solutions (4) and (4') agree with the central experimental value (1) to within a remarkable accuracy $\sim\!+2 \times 10^{-10}$, i.e. $\sim\!0.2$ ppb or $\sim\!+0.3$ S.D. This accurate agreement supports the notion of a close relation between the fine structure constant at zero momentum transfer and the new universal physical constant $\alpha_o$. Future comparison of the solutions (4) and (4') with improved new experimental $\alpha$-data in prospect [2, 1] should be suggestive.

**3. Relation to electroweak theory.** The electroweak theory (EWT) [6] as part of the Standard Model is proven a highly successful precise description of all known lepton physics interaction phenomena except gravity. The values of the two interaction constants[1] $\alpha$ and $\alpha_W$ and particle flavor quantities such as mass ratios and mixings remain free parameters in the EWT taken from the experimental data. Therefore, there is no incongruity between tracing the values of EWT free parameters from the new physical constant $\alpha_o$ via relations suggested by the experimental data. The precise Eq.(3) for the fine structure constant may replace the anthropic principle reasoning for justifying the concrete choice of the free parameter $\alpha_{Data}$ by the experimental data. Eq.(3) is twofold interesting: 1) *in principle,* as a likely answer to the question of where does the absolute fine structure constant numerical value $\alpha_{Data}$ come from, and 2) *in* context of a system of other relevant experimental indications in favor of the introduced new flavor-electroweak constant $\alpha_o$ [4]. Indeed, as a remarkable fact the constant $\alpha_o$ simultaneously generates the values of both known basic electroweak interaction constants $\alpha$ and $\alpha_W$ at the different pole $Q^2$-values of the photon and W-boson propagators. While the precise connection between $\alpha(Q^2 = 0)$ and $\alpha_o$ is given by (3), the accurate value of the second electroweak interaction constant $\alpha_W$ is given in terms of $\alpha_o$ by the relation $\alpha_W \equiv \alpha_W(Q^2 = M_W^2) \cong \alpha_o \log \alpha_o^{-1} = 5e^{-5}$. If parameterized as $\alpha_W = 5\exp[-5(1+\Delta)]$, it follows from the experimental data that $|\Delta| < 0.001$, comp. [7] – it is a hint from the data on unification of the interaction constants $\alpha_W$ and $\alpha$. Note that these facts are displayed against a background of experimental evidence that the constant $\alpha_o$ is a main factor in the description of the bare mass ratios of charged leptons, and probably also neutrinos and quarks [5].

---

[1] $\alpha_W = \alpha/\sin^2\theta_W$ is the dimensionless weak interaction constant and $\theta_W$ is the Weinberg weak mixing angle.

**4. On the structure of Eq.(3).** I should comment on the heuristic origin of the accurate equation (3) for the fine structure constant $\alpha$. In EWT the constant $\alpha \equiv \alpha(Q^2 = 0)$ is determined only by the experimental value $\alpha_{exp}$. This status of the fine structure constant $\alpha$ as a free parameter in the theory is in essence not changed by the Eq.(3), which determines $\alpha$ by the new universal constant $\alpha_o$ instead of $\alpha_{exp}$. Since the constant $\alpha$ experimentally appears close to $\alpha_o$ (~8%), the main term of the assumed connection between $\alpha$ and $\alpha_o$ in (3) follows from the observation that it is a nonlinear dressing of $\alpha$ by the natural here close to unity exponential factors $(\exp \alpha)^n$, **n** = 0,1,2. A few steps of selection, especially simple in logarithmic form, lead to the nonlinear equation

$$\exp 2\alpha \, \log(\exp\alpha/\alpha) \cong \log(1/\alpha_o) = 5 \qquad (5)$$

with the solution

$$\alpha \cong 1/137.0383, \quad (\alpha - \alpha_{Data})/\alpha_{Data} \cong -1.7 \times 10^{-5}. \qquad (6)$$

If future precise cosmological observations and computer simuulations will discover a value $\alpha < \alpha_{Data}$ comparable to the estimation (6) for the fine structure constant $\alpha = \alpha(Q^2 = 0)$ at some early time of our universe, the cosmological meaning of Eq.(3) may be establish, comp. ref.[8].

The difference between the right and left sides (at $\alpha = \alpha_{Data}$) in the obtained after exponentiation of (5) equation (5'),

$$(\exp\alpha/\alpha)^{\exp 2\alpha} = 1/\alpha_o, \qquad (5')$$

appears equal to $(\alpha/\pi)$ to within an accuracy of $\sim 2\times 10^{-3}$, more accurately $\sim (\alpha_o/\pi)$. It is a special *decisive hint* from the precise experimental data [2] at the linear in $\alpha$ term of equation (3). So, Eq.(3) for the fine structure constant $\alpha \equiv \alpha(Q^2 = 0)$ is definitely generated by: 1) the new primary physical constant $\alpha_o$, 2) hints from the new experimental data [2] and 3) new electroweak calculations [1, 3].

**5. Conclusions**. To conclude, a system of observations on lepton mass and electroweak experimental data indicates [4, 5] that $\alpha_o$ is likely a new universal basic dimensionless physical constant uniting lepton flavor and electroweak phenomenologies. In the present Note, this new constant $\alpha_o$ as a source generates the precise equation (3) for the fine structure constant $\alpha$ and its solution (4) which is a match for the most accurate to date experimental value $\alpha_{exp}$ (1).

I would like to thank M. Passera for interest in the match and information.